# Liquid Argon Cryogenic Detector Calibration by Inelastic Scattering of Neutrons

**Sergey Polosatkin,**[*a,b,c] **Evgeny Grishnyaev,**[a] **and Alexander Dolgov**[b]

[a] *Budker Institute of Nuclear Physics SB RAS,*
   *630090, Lavrent'eva av. 11, Novosibirsk, Russia*
[b] *Novosibirsk State University,*
   *630090, Pirogova st. 1, Novosibirsk, Russia*
[c] *Novosibirsk State Technical University,*
   *630073, Karl Marx av. 20, Novosibirsk, Russia*

   *E-mail*: S.V.Polosatkin@inp.nsk.su

ABSTRACT: A method for calibration of cryogenic liquid argon detector response to recoils with certain energy – 8.2 keV - is proposed. The method utilizes a process of inelastic scattering of monoenergetic neutrons produced by fusion DD neutron generator. Features of kinematics of inelastic scattering cause sufficient (forty times) increase in count rate of useful events relative to traditional scheme exploited elastic scattering with the same recoil energy and compatible energy resolution. The benefits of the proposed scheme of calibration most well implemented with the use of tagged neutron generator as a neutron source that allows to eliminate background originated from casual coincidence of signals on cryogenic detector and additional detector of scattered neutrons.

---

[*] Corresponding author.

**Contents**

**1. Introduction**

Two-phase noble gas cryogenic detectors are a new class of instrument for particle physics, cosmology, and medicine. An operation principle of the instrument is based on the measurements of free electrons, scintillation light, or heat produced by a particle passing through liquid noble gas serving as active medium of the detector. High sensitivity and low noise of such detectors make it possible to use them for detection of rare events with small energy release like coherent neutrino scattering or Dark Matter (DM) detection. Liquid argon is considered to be the most suitable medium for these purposes. The technique of cryogenic detectors is actively developing and several large scale experiments are started [1-5].

Scattering events of both DM particles and neutrinos produce recoiling nuclei that are slowed down in the detector medium and cause its ionization and excitation. Measurements of recoil spectra require calibration of the detector that is determination of dependence between recoil energy and detector response. Note that instead of electronic recoils produced by beta- or gamma radiation this dependence is substantially nonlinear because of secondary nuclear energy losses and several other effects. These nonlinear effects often accounted by the use of a quenching factor that is the ratio of responses from nucleus and electronic recoils with the same energy. Therefore correct calibration procedure should consist on the measurements of response from nuclear recoils with known energy. The energy regions of interest corresponds to estimated energies of recoils appeared in the explored process and lies in the ranges of 0.1-1 keV for neutrinos and 1-50 keV for DM scattering [6, 7]. In the presented paper we propose new method of producing recoiling argon nuclei with certain energy that can be used for calibration of WIMP detectors.

**2. LAr detector calibration schemes**

Elastic scattering of neutrons is a process commonly used for the detectors calibration. There are two main approaches for this calibration. The first utilizes neutron sources with wide spectrum such as $^{252}$Cf or Pu-Be. Quenching factor is reconstructed by fitting of experimental pulse height spectrum to Monte-Carlo simulations of neutron scattering. Actually reconstruction of quenching factor represents solving of inverse problem so it requires detailed information about spectrum of neutrons and accuracy of such measurements is rather low. Another approach uses a monoenergetic neutron source. The general scheme of the calibration in this case is the follows (fig.1a): a flux of monoenergetic neutrons with low angular spread is directed to the detector. Neutrons scattered at a certain angle are counted by additional detector of scattered neutrons (DSN). Scattering events are recognized by coincidence between the main detector and the DSN pulses. Recoil energy for detected event can be derived from scattering kinematics using known values of primary neutron energy and scattering angle

The energy of the recoils is equal to

$$E_{rec} = 2 \cdot E_0 \frac{mM}{(m+M)^2} \left( 1 - \cos\theta \sqrt{1 - \left(\frac{m}{M}\sin\theta\right)^2} - \frac{m}{M}\sin^2\theta \right),$$

where En – energy of incident neutron, M – mass of recoiling nucleus, m- mass of neutron, q-scattering angle. Approximately

$$E_{rec} = 2 \cdot E_0 \frac{mM}{(m+M)^2} (1 - \cos\theta)$$

Unfortunately there are only a few types of neutron sources that generate monoenergetic neutrons. Usually DD neutron generators producing neutrons with the energy 2.45 MeV are used for detector calibration. Such neutrons produce argon recoils with energies from 0 to 233 keV.

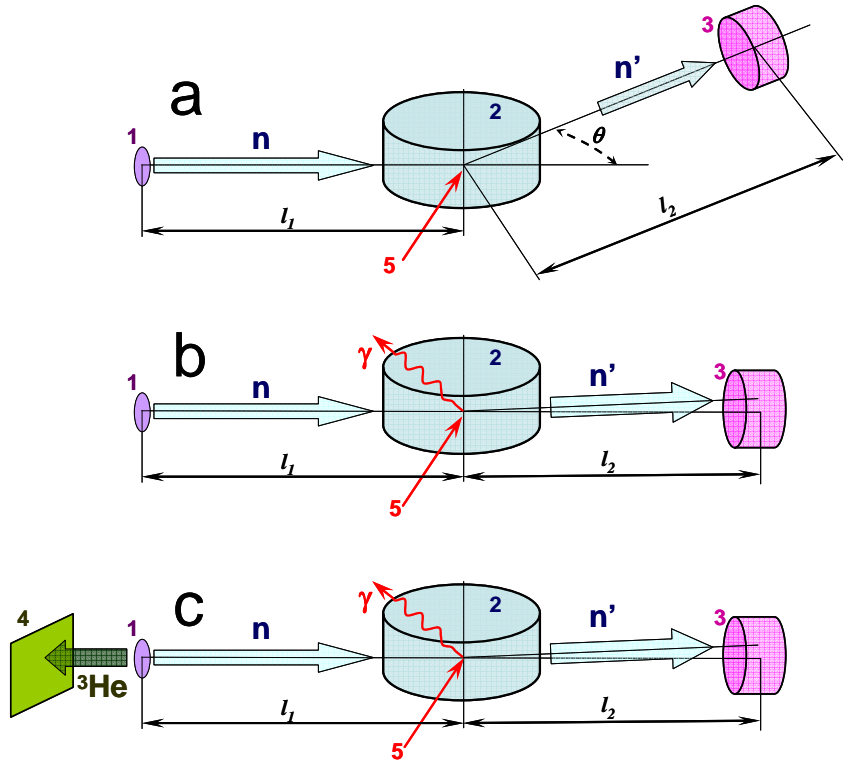

*Fig.1 Schemes of cryogenic detector calibration by monoenergetic neutron source, a- calibration with elastic scattering, b – calibration with inelastic scattering, c – calibration with inelastic scattering with a tagged neutron generator; 1 – neutron source, 2 – cryogenic detector, 3 – scintillation detector of scattered neutrons (DSN), 4 – build-in detector of $^3$He nuclei, 5 – scattering point*

Actually the lower limit of energy range of recoils that can be applied for calibration is restricted by "geometric" errors caused by uncertainty of scattering angle due to finite dimensions of the detectors. The uncertainty of recoil energy approximately equals to

$$\frac{\Delta E_{rec}}{E_{rec}} \approx \frac{\sin\theta}{1-\cos\theta}\Delta\theta$$

where θ - scattering angle. This error rapidly rises with decreasing of scattering angle (that is recoil energy). For recoil energies 100, 10, and 1 keV (scattering angles 80.4, 23.3, and 7.3 degrees) the geometric errors are equal to 1.1Δθ, 4.7Δθ и 15Δθ. In this regard practically the recoil energy range that can be covered by calibration via neutron elastic scattering is restricted by energy about 10 keV [8]. Hence it is important to develop alternative calibration schemes suitable for recoils with energy below 10 keV.

Here we propose a new calibration scheme based on the process on inelastic scattering of neutrons (fig.1b). Collisions of neutrons with nuclei of noble gas filling the detector may cause excitation of the nuclei with following radiative decay. For $^{40}$Ar first exciting level has the energy 1.46 MeV. Hence inelastic collision may be expressed as

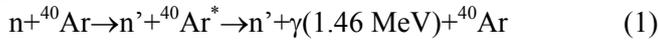

n+$^{40}$Ar→n'+$^{40}$Ar$^*$→n'+γ(1.46 MeV)+$^{40}$Ar    (1)

A feature of kinematic of inelastic scattering is that the energy of recoil tends to certain non-zero value with decreasing of scattering angle. This value is equal to 8.2 keV for the mentioned process (1) and energy of primary neutrons 2.45 MeV. For low-angle scattering energy of recoils weakly depends on scattering angle, accordingly probability of generation of recoils with the energy near 8.2 keV in inelastic scattering sufficiently exceeds corresponding probability for elastic scattering. Thereby registration of recoils originated in low-angle inelastic scattering of neutrons allows to implement calibration of LAr detector for certain energies with increased (as compared to standard scheme using elastic scattering) efficiency.

## 3. Inelastic scattering of neutrons

Energy of excited argon recoil born in the inelastic collisions can be found as:

$$E^*_{rec} = 2\cdot E_0 \frac{mM}{(m+M)^2}\left(1-\cos\theta\sqrt{1-\left(\frac{m}{M}\sin\theta\right)^2 - \frac{\varepsilon}{E_0}\left(1+\frac{m}{M}\right)} - \frac{m}{M}\sin^2\theta - \frac{1}{2}\frac{\varepsilon}{E_0}\left(1+\frac{m}{M}\right)\right)$$

or approximately:

$$E^*_{rec} = 2\cdot E_0 \frac{mM}{(m+M)^2}\left(1-\cos\theta\sqrt{1-\frac{\varepsilon}{E_0}} - \frac{1}{2}\frac{\varepsilon}{E_0}\right)$$

where ε - energy of excitation of the nuclear level.

In the following we will summarize and compare cross-sections of processes of elastic scattering and inelastic scattering via first exciting level for collisions on neutrons with $^{40}$Ar nucleus for neutrons with energies 2.45 and 14.1 MeV, that can be produced by DD and DT neutron generators.

First excited level of $^{40}$Ar has the energy ε=1.46 MeV. Energies of recoils vs. scattering angle for processes of interest are shown on the fig.2. For low-angle inelastic scattering these energies tend to constant values 8.2 and 0.95 keV for energies of primary neutrons 2.45 and 14.1 MeV.

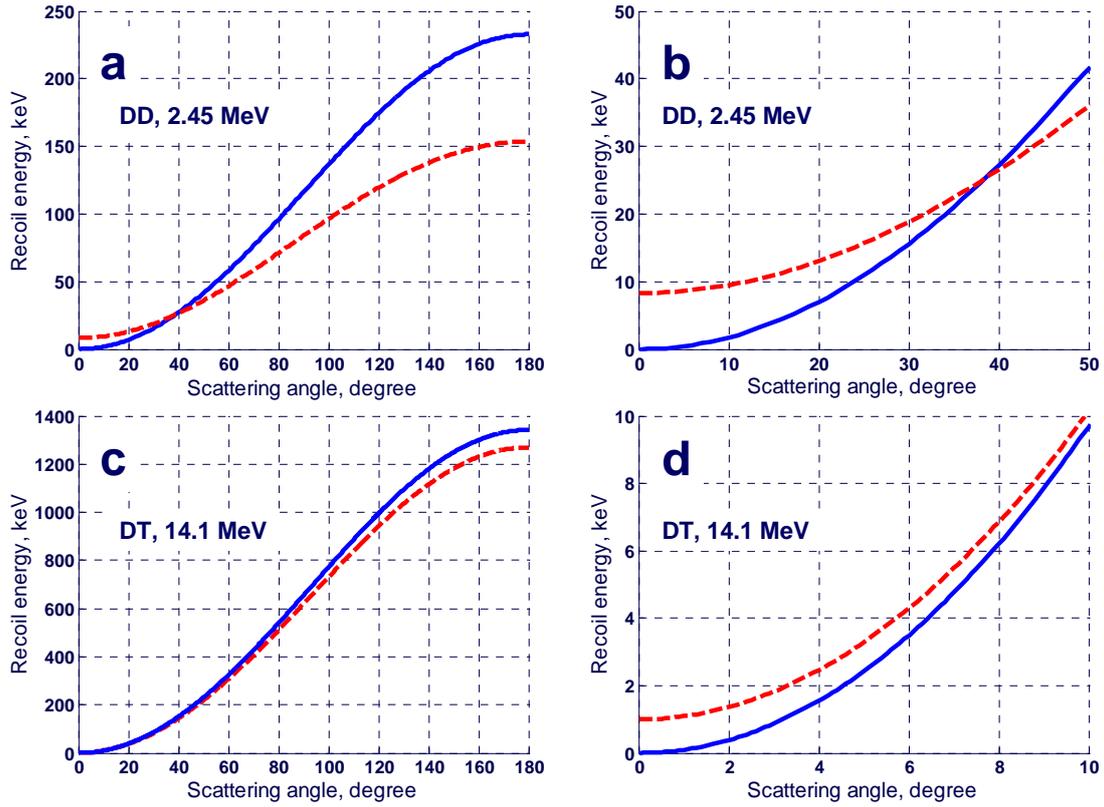

*Fig.2 Dependence of Ar recoils energy on neutron scattering angle, a,b- DD neutrons (2.45 MeV), c,d – DT neutrons (14.1 MeV); solid lines – elastic scattering, dashed lines – inelastic scattering via first exciting level (1.46 MeV)*

Differential cross-sections for low-angle inelastic scattering of DD and DT neutrons are equal to 50 mb/str. Equivalent cross-sections for elastic scattering with production of recoils with the same energy (that is scattering on an angle 22 deg for DD neutrons and on 3 deg for DT neutrons) are 0.8 and 1.8 b/str [9].

The excited recoils radiativelly decay to ground state with emission of 1.46 MeV photon. The timescale of the decay is on the order of $10^{-12}$s, an angular distribution of photon emission is close to uniform with accuracy level ~10% [10]. Since the photon carries away a momentum ε/c, final energy of argon recoil is equal to

$$E_{rec} = E_{rec}^* + \frac{1}{2}\frac{\varepsilon^2}{Mc^2} - \sqrt{\frac{2E_{rec}^*}{Mc^2}} \cdot \varepsilon \cdot \cos\psi,$$

where ψ – an angle between directions of excited recoil velocity and photon emission.

Notice that photon emission causes dispersion of energies of recoils with certain scattering angle. That is in the discussed process in contrast to elastic scattering there are no unambiguous relation between neutron scattering angle and final recoil energy. For low-angle inelastic scattering the energy distribution functions of recoils is close to square-wave with edge energies 7.3-9.2 keV for DD and 0.65-1.3 keV for DT neutrons.

## 4. Simulation of neutron scattering

Inelastic scattering of neutrons (the process (2)) can be applied for calibration of liquid argon detectors of weakly interacting particles on the two certain energies of recoils - 8.28 and 0.98 keV. Advantage of exploiting of this process is low sensitivity to geometric errors originated from finite spatial resolution of detectors that allows to achieve respectively high count rate of scattering events. Actually the count rate in the coincidence scheme shown on the fig.1 is proportional to fourth power of geometric dimensions of detectors, so increase of size of detector without loss of accuracy of recoil energy determination leads to sufficient increase in count rate of the scheme. Drawbacks of the use of inelastic scattering are low cross-section of this process (more then twenty time less than elastic scattering) and inherent dispersion of recoil energy that restrict accuracy of the calibration.

We produced simulation of neutron scattering in the LAr calibration scheme for quantitative comparison of exploiting of elastic and inelastic scattering. The simulation was produced for the detectors available now in the Budker Institute of Nuclear Physics. A two-phase cryogenic argon detector with GEM amplification and silicon photomultiplier array read-out has cylindrical sensitive volume with dimensions D50x50 and spatial resolution in the horizontal plane 10 mm. Stilbene scintillator with dimensions D40x40 mm can be applied as a secondary detector of scattered neutrons. Distances between neutron source and detectors ("levers" of calibration scheme) are considered as free parameters of modeling.

There are several universal sophisticated codes for Monte-Carlo simulation of neutron interaction with matter such as MCNP [11], GEANT4 [12], and FLUKA [13]. These codes in principle can be applied for the modeling of neutron scattering. At the same time the codes have excessive complexity for considered task and, on the other hand, require developing of additional modules for extraction of recoil spectra. Hence custom-made code Scattronix [14] was developed and applied for simulation. Several assumptions allow to simplify code and increase its performance. First, both multiple scattering of neutrons in the LAr detector and energy release from 1.46 MeV gamma produced in inelastic scattering are disregarded. The reason is that the size of the detector sufficiently less then neutron free path (that exceeds 10 cm in whole region of expectable energies of neutrons) and attenuation length of 1.46 MeV gammas equals to 15 cm [15]. Second, only the processes (1) and (2) are considered in the simulation. This assumption really correct for 2.45 MeV DD neutrons for which these two processes dominate over all another. For 14.1 MeV DT neutrons the situation is more complex because cross-sections of (n,2n) reaction and inelastic scattering via upper levels are compatible with elastic scattering cross-section. In the Scattronix code these processes are considered as "disappearing" of primary neutron without including in statistics recoils produced in these reactions. Accordingly Scattronix code accurately accounts exact geometry of the system but doesn't allow modeling of background originated form neglected processes and neutron scattering in the detector environment.

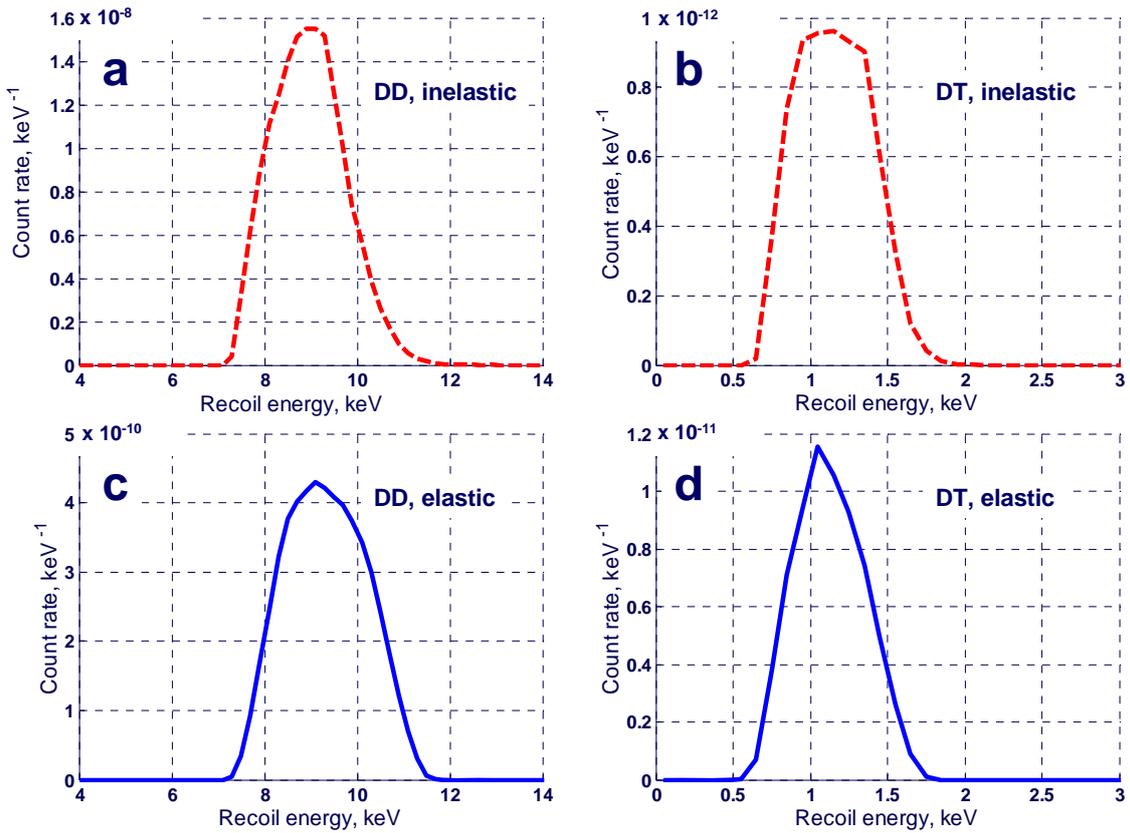

*Fig.3 Count rates of DSN coincidence schemes (Scattronix simulations); **a** – inelastic scattering of DD neutrons, $E_0=2.45$ MeV, $\theta=0°$, $l_1=l_2=30$ cm, **b** – inelastic scattering of DT neutrons, $E_0=14.1$ MeV, $\theta=0°$, $l_1=l_2=145$ cm, **c** – elastic scattering of DD neutrons, $E_0=2.45$ MeV, $\theta=22.5°$, $l_1=l_2=200$ cm, **d** - elastic scattering of DT neutrons, $E_0=14.1$ MeV, $\theta=3.2°$, $l_1=l_2=470$ cm; count rates normalized to full neutron flux produced by neutron generator*

Pulse height spectra discriminated by coincidence scheme for DD and DT neutrons and different geometries of the system are presented on the fig.3. Fig.3a and 3b corresponds to inelastic scattering of DD and DT neutrons to low angle with levers of the calibration scheme 30 and 145 cm. Fig.3c and 3d represent pulse-height spectra of elastic scattering of DD and DT neutrons with scattering angles 22.5 and 3.2 degrees and lever lengths 200 and 470 cm (the count rates normalized to full neutron flux of neutron generator). These geometries provide similar energy resolution in elastic and inelastic scattering so they allow quantitative comparison of count rates of different schemes. Fig.3 shows that the use of inelastic scattering of DD neutrons for detector calibration on the recoil energy near 8.5 keV allows 40-fold increase of count rate of useful events. At the same time for DT neutrons (recoil energies near 1 keV) inelastic scattering is less favorable than the elastic.

## 5. Background discrimination in inelastic calibration scheme

The main problem of practical realization of inelastic scattering calibration scheme is impossibility to shield the detector of scattered neutrons from a primary neutron flux. Accordingly nonideality of coincidence scheme would cause appearance of background on the pulse height spectra due to casual coincidence of LAr and DSN detector pulses originated by two different neutrons. A level of such background depends on neutron flux value and temporal resolution of coincidence scheme. Taken reasonable values – the generator neutron yield $10^6$ s$^{-1}$ and temporal resolution 10 μs (determined by drift velocity of electrons in liquid argon) – we will estimate that background level is close to count rate of scattering events. Estimated for mentioned conditions coincidence spectrum is shown on the fig.4.

The sharp peak near 9 keV represents "useful" events of inelastic scattering. The peak near zero energy corresponds to elastic scattering of neutrons, wide background is originated from casual coincidence of pulses in the two detectors.

Due to this casual coincidence background more then ten events are required for the inelastic scattering peak recognition on the significance level 1%, that, in some extent, eliminates benefits of the discussed scheme of calibration. The problem of casual coincidence background can be overcome by the use of a neutron generator with a neutron tagging system (associated particle neutron generator [16,17]) (Fig.1c). In a generator of such type build-in particle detector records signals from fast helium nucleus that accompany neutron origination in the reaction

D+D$\rightarrow$ n (2.45 MeV) + 3He (0.82 MeV)

This signal gives nanosecond accuracy reference for every neutron born in the generator that sufficiently increases temporal resolution of the coincidence scheme. Moreover since neutrons looses sufficient energy in the inelastic scattering process the neutrons after inelastic scattering can be distinguished from direct and elastically scattered neutron by time-of-flight analysis. Actually for calibration scheme with 30 cm levers direct neutron flight time between neutron source and DSN detector is 28 ns whereas flight time for inelastically scattered neutrons is equal to 36 ns. Consequently the use of tagged neutron generator allows to eliminate background in the noble-gas detector calibration that is especially important for realization of suggested inelastic calibration scheme.

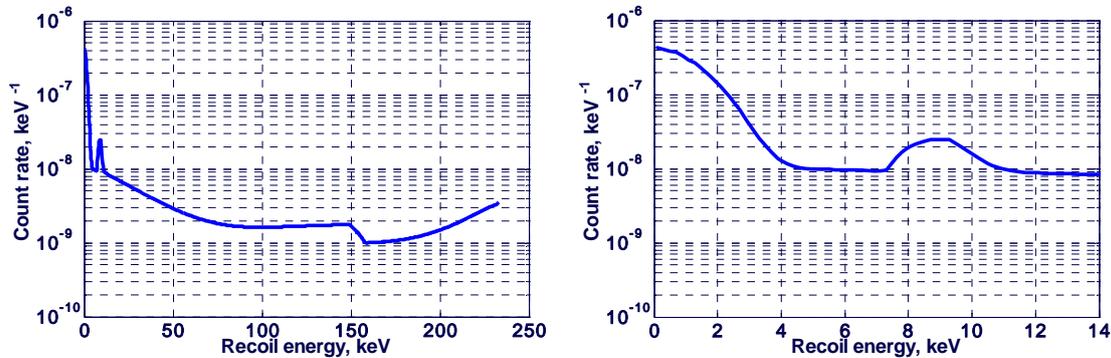

*Fig.4 Pulse-height spectrum of coincidence scheme with near-axis DSN detector (Scattronix simulation), $E_0$=2.45 MeV, $\theta$=0°, $l_1$=$l_2$=30 cm, primary neutron flux $10^6$ s$^{-1}$, temporal resolution of coincidence scheme 10 μs.*

## 6. Conclusion

Low-angle inelastic scattering of monoenergetic neutrons allows to calibrate response of cryogenic noble gas detectors to recoils with certain energy that depends on the energy of primary neutrons and recoil mass and nuclear structure. Specific energy of recoils that can be produced by DD neutrons (2.45 MeV) in argon is 8.2 keV. Features of kinematics of inelastic scattering cause sufficient (forty times) increase in count rate of useful events relative to traditional scheme exploiting elastic scattering with the same recoil energy and compatible energy resolution. The benefits of the proposed scheme of calibration most well implemented with the use of tagged neutron generator as a neutron source that allows elimination of background originated from casual coincidence of signals on cryogenic detector and additional detector of scattered neutrons.

## Acknowledgments

This work was financially supported by the Ministry of Education and Science of the Russian Federation and by the Russian Government, Project no. 11.G34.31.0047.